\documentclass[aps,preprint]{revtex4}
\usepackage{amsfonts}
\usepackage{amsmath}
\usepackage{amssymb}
\usepackage{graphicx}

\begin{document}

\title{Relativistic Dissipative Hydrodynamics: A Minimal Causal Theory}
\author{T. Koide, G. S. Denicol, Ph. Mota and T. Kodama}
\affiliation{Instituto de F\'{\i}sica, Universidade Federal do Rio de Janeiro, C. P.
68528, 21945-970, Rio de Janeiro, Brazil}
\keywords{Relativistic Hydrodynamics, Dissipation, Memory effect, Causality}
\pacs{47.10.-g,25.75.-q}

\begin{abstract}
We present a new formalism for the theory of relativistic dissipative
hydrodynamics. Here, we look for the minimal structure of such a theory
which satisfies the covariance and causality by introducing the memory
effect in irreversible currents. Our theory has a much simpler structure and
thus has several advantages for practical purposes compared to the
Israel-Stewart theory (IS). It can readily be applied to the full
three-dimensional hydrodynamical calculations. We apply our formalism to the
Bjorken model and the results are shown to be analogous to the IS.
\end{abstract}

\volumeyear{year}
\maketitle

\section{Introduction}

The ideal hydrodynamical description for the dynamics of hot and dense
matter achieved in RHIC experiments works amazingly well, particularly for
the behavior of collective flow parameters. Together with other signals, the
success of the approach is considered as the indication of the emergence of
a new state of strongly interacting matter, the plasma of quarks and gluons
(QGP). The comparison between RHIC and SPS results shows that this new state
of matter is formed at the very early stage of the relativistic heavy ion
collisions for RHIC energies. The nature of the QGP seems rather a strongly
interacting fluid (sQGP) than a ideal free parton gas \cite{Shuryak-ideal}
which still flows like an ideal-fluid. This idea, that the QGP behaves as
real ideal fluid, raised an interesting perspective, since the viscosity for
the strong coupling limit of 4D conformal theory obtained from the
supersymmetric string theory in an (Anti de Sitter) AdS space found to be
very small \cite{Son-viscosity}. On the other hand, Hirano and Gyulassy
argue that this is due to the entropy density of the QGP which is much
larger than that of the hadronic phase \cite{Miklos}.

At any rate, we know that there still exist several open problems in the
interpretation of data in terms of the hydrodynamical model \cite{OpenProb}.
These questions require careful examination to extract quantitative and
precise information on the properties of QGP. In particular, we should study
the effect of dissipative processes on the collective flow variables.
Several works have been done in this direction \cite{Viscosity}. However,
strictly speaking, a quantitative and consistent analysis of the viscosity
within the framework of relativistic hydrodynamics has not yet been done
completely. This is because the introduction of dissipative phenomena in
relativistic hydrodynamics casts difficult problems, both conceptual and
technical. Initially Eckart, and later, Landau-Lifshitz introduced the
dissipative effects in relativistic hydrodynamics in a covariant manner \cite%
{Eckart,LL}. It is, however, known that their formalism leads to the problem
of acausality, that is, a pulse signal propagates with infinite speed. Thus,
relativistic covariance is not a sufficient condition for a consistent
relativistic dissipative dynamics \cite{MullerReviw,Jou}.

To cure this problem, relativistic hydrodynamics in the framework of
extended thermodynamics was developed by M\"{u}ller \cite{mueller} and later
by Israel and Stewart \cite{Israel,Israel2,Israel3}. From the kinetic point
of view \cite{Grad}, this formalism corresponds to the extension of
equilibrium thermodynamics to include the second order moments of kinetic
variables. This is the reason why this theory is usually referred to as the
second order theory. However, this theory is too general and complex,
introducing many unknown parameters from the point of the present QCD
dynamics. Moreover, the irreversible currents are treated as independent
variables in addition to the usual hydrodynamical degrees of freedom (the
velocity field plus densities of extensive quantities) and should be
obtained by solving highly coupled partial differential equations. These
features make the application of the theory to practical problems very
difficult.

On the other hand, even if we overcome the technical difficulties to apply
the complete theory to the analysis of relativistic heavy ion reactions, it
is very unlikely that we can extract such a detailed information from the
experimental data at present moment. This is because there exist many
uncertainties of the hydrodynamical approach itself such as the initial
condition, finite size correction, event-by-event fluctuations, particle
decoupling process etc. \cite{OpenProb}. 
However, there exists some
interesting questions which requires a correct treatment of the viscosity in
a realistic 3D simulations, such as a possible generation of shock wave and
its propagation in the QGP triggered by a jet\cite{Shock-Horst}. Analysis of
such phenomena will certainly give important information of new, genuine
hydrodynamical properties of the QGP. Therefore, what we need now is 
not a very general theory of relativistic dissipative hydrodynamics
with so many unknown parameters but one with the minimum number of
parameters necessary to preserve fundamental principles such as causality
and entropy production.

As a matter of fact, it is known that the Israel-Stewart theory (IS) is not
the unique approach to relativistic dissipative hydrodynamics. To authors'
knowledge, there is at least one other causal theory called the divergence
type \cite{Div1,Div2,Div3,Div4,Div5}. In this work, we present a very simple
alternative theory which satisfies the minimal conditions mentioned above.
We show that the causality problem of the Landau-Lifshitz formalism can be
solved by introducing a memory effect. This memory effect is characterized
by the relaxation time $\tau_{R}$, so that our theory introduces only one
additional parameter to the usual bulk viscosity, shear viscosity and
thermal conduction coefficients of the Navier-Stokes equation. Our theory
recovers the relativistic Navier-Stokes equation in the limit of vanishing
values of this relaxation time.

As described later more in detail, our approach has a fundamental advantage
from the practical point of view in addition to its physical simplicity. The
dissipative terms are explicitly given by the integral of the independent
variables of the usual ideal hydrodynamics. Thus, the implementation of our
method to the existing ideal hydro-codes is straightforward, particularly,
to those based on the local Lagrangian coordinate system such as SPheRIO 
\cite{Spherio,Review}.

The present paper is organized as follows. In the next section, we briefly
review the problem of the acausal propagation in the diffusion equation and
the method to cure this problem in terms of the memory effect, which leads
to the so-called telegraphist's equation. In the Sec. III, we analyse the
structure of the Landau-Lifshitz theory of relativistic dissipative
hydrodynamics and introduce the memory effect to solve the acausal problem
due to its parabolic nature. We thus obtain the dissipative hydrodynamical
equations with the minimum number of parameters which satisfies causality.
In Sec. IV, we discuss the problem of entropy production in our formalism.
In Sec.V, we apply our equation to the Bjorken solution, and compare with
the previous analyses \cite{Muronga1,Muronga2,Muronga3, Baier1,Baier2}. In
Sec. VI, we summarize our results and discuss possible immediate
applications.

\section{Diffusion Equation and Acausality}

The fundamental problem of the first order theory like the Navier-Stokes
theory comes from the fact that the diffusion equation is parabolic. This
means that the velocity of signal propagation is infinite. This problem was
first addressed by Cattaneo \cite{Cattaneo} in the case of heat conduction.
He discussed that the problem of acausal propagation in usual diffusion
equation, 
\begin{equation}
\frac{\partial}{\partial t}n=\zeta\nabla^{2}n,  \label{Diffusion}
\end{equation}
can be cured by the introduction of the second order time derivative as 
\begin{equation}
\tau_{R}\frac{\partial^{2}n}{\partial t^{2}}+\frac{\partial n}{\partial t}%
=\zeta\nabla^{2}n,  \label{tele}
\end{equation}
thus converting the parabolic equation to a hyperbolic one. 
Here, $\zeta$ and $\tau_R$ are a diffusion constant and a relaxation time,
respectively. Eq.(\ref{tele}) is sometimes called as telegraphist's
equation. For a suitable choice of the parameters $\tau_{R}$ and $\zeta$, we
can recover the causal propagation of diffusion process. In fact, the
maximum velocity of the signal propagation governed by this equation is
shown to be \cite{MorseFeshbach} 
\begin{equation}
v_{\max}=\sqrt{\frac{\zeta}{\tau_{R}}}.  \label{vmax}
\end{equation}
To be consistent with the special relativity, we should have 
\begin{equation}
\tau_{R}\geq\frac{\zeta}{c^{2}}.  \label{tmin}
\end{equation}
The diffusion equation, Eq.(\ref{Diffusion}), corresponds to $\tau_{R}=0$
and hence $v_{\max}\rightarrow\infty.$

The physical origin of above telegraphist's equation can be understood as
follows. The diffusion process is a typical relaxation process for a
conserved quantity. Thus, it should satisfy the equation of continuity, 
\begin{equation}
\frac{\partial n}{\partial t}+\nabla\cdot\vec{j}=0.  \label{Cont}
\end{equation}
In non-equilibrium thermodynamics, the irreversible current $\vec{j}$ is
assumed to be proportional to a thermodynamic force, 
\begin{equation}
\vec{j}=-L\vec{F},  \label{Curr}
\end{equation}
where the Onsager coefficient $L$ is, in general, a function of
thermodynamic quantities. Here, we assume it to be constant for the sake of
illustration. When $n$ is a number density, the thermodynamic force is given
by the gradient of $n$ because of Fick's law, 
\begin{equation}
\vec{F}=\nabla n.  \label{Fick}
\end{equation}
Substituting Eq.(\ref{Curr}) into Eq.(\ref{Cont}), we get the diffusion
equation (\ref{Diffusion}).

Fick's law tells us that the above diffusion process is induced by
inhomogeneous distribution. In Eq.(\ref{Curr}), the space inhomogeneity
immediately gives rise to the irreversible current. However, this is a very
idealized situation. In general, the generation of irreversible currents has
a time delay. Thus, we may think of memory effects within the linear
response of the system. Phenomenologically, this can be done by introducing
the following memory function \cite{Koide2}, 
\begin{align}
G\left( t,t^{\prime}\right) & =\frac{1}{\tau_{R}}e^{-\left( t-t^{\prime
}\right) /\tau_{R}},\ t\geq t^{\prime}  \label{relax} \\
& =0,\ \ \ \ \ \ \ \ \ \ \ \ \ \ \ \ \ t<t^{\prime}
\end{align}
where $\tau_{R}$ is a constant (relaxation time) and rewrite Eq.(\ref{Curr})
as 
\begin{equation}
\vec{j}=-\int_{-\infty}^{t}G\left( t,t^{\prime}\right) L\vec{F}\left(
t^{\prime}\right) dt^{\prime}.  \label{retard}
\end{equation}
In the limit of $\tau_{R}\rightarrow0,$ we have $G\left( t,t^{\prime}\right)
\rightarrow\delta\left( t-t^{\prime}\right) $ so that the original equation (%
\ref{Curr}) is recovered \cite{supernova}. The time derivative of the
irreversible current (\ref{retard}) leads to 
\begin{equation}
\frac{\partial\vec{j}}{\partial t}=-\frac{1}{\tau_{R}}L\vec{F}\left(
t\right) -\frac{1}{\tau_{R}}\vec{j}.
\end{equation}
This is called the Maxwell-Cattaneo type equation. Substituting into the
equation of continuity (\ref{Cont}), we arrive at 
\begin{align*}
\frac{\partial^{2}n}{\partial t^{2}} & =\nabla\cdot\left( \frac{1}{\tau _{R}}%
\zeta\vec{F}\left( t\right) +\frac{1}{\tau_{R}}\vec{j}\right) \\
& =\frac{1}{\tau_{R}}\left( -\frac{\partial n}{\partial t}+\zeta\nabla
^{2}n\right) ,
\end{align*}
which is exactly Eq. (\ref{tele}).

We here showed the problem of acausality in the diffusion equation, which is
based on non-equilibrium thermodynamics (the first order theory), and how
this can be solved by introducing the memory function. For more microscopic
derivation of telegraphist's equation and foundation of the second order
theory, see Refs. \cite{Jou,Koide1,Koide2,Gavin}. In particular, it was
recently shown that the macroscopic equation of motion, which is obtained by
using systematic coarse-grainings from the Heisenberg equation of motion, is
not the diffusion equation but telegraphist's equation \cite{Koide1}.

Till now, we have emphasized the importance of the memory effect in the view
of causality, but there exists another important reason to introduce the
memory effect. It was pointed out that the diffusion equation also
contradicts with sum rules associated with conservation laws. This can also
be solved by replacing the diffusion equation with telegraphist's equation 
\cite{Koide1}. Relativistic dissipative hydrodynamics should be consistent
with QCD, where there are conserved quantities and associated sum rules. In
this sense, the presence of memory effect should be inevitable.

\section{Minimal Causal Hydrodynamics}

Eckart and Landau-Lifshitz derived the relativistic dissipative
hydrodynamics following non-equilibrium thermodynamics as discussed in the
preceding section \cite{Eckart,LL}. Their theories are just the covariant
versions of the Navier-Stokes equation and the corresponding equations still
continue to be parabolic. Therefore, they do not satisfy causality and some
modification should be required. In the IS and the divergence type theory,
the definition of the entropy four-flux is generalized and, to satisfy the
second law of thermodynamics, modified thermodynamic forces are obtained. In
this section, we propose another approach, where the problem of acausality
is solved by introducing the memory effect like Eq. (\ref{retard}).

For this purpose, let us first analyse briefly the structure of the
Landau-Lifshitz theory (LL). The hydrodynamical equation of motion is
written as the conservation of the energy-momentum tensor , 
\begin{equation}
\partial_{\mu}T^{\mu\nu}=0,  \label{divTmunu}
\end{equation}
together with the conservation of a quantity, for example, the baryon
number, 
\begin{equation}
\partial_{\mu}N^{\mu}=0.  \label{divN}
\end{equation}
In the LL, it is assumed that thermodynamic relations are valid in the local
rest frame of the energy-momentum tensor. The energy-momentum tensor is
expressed as 
\begin{equation}
T^{\mu\nu}=\varepsilon u^{\mu}u^{\nu}-P^{\mu\nu}\left( p + \Pi\right)
+\pi^{\mu\nu},  \label{NeTmunu}
\end{equation}
where, $\varepsilon$, $p$, $u^{\mu},$ $\Pi$ and $\pi^{\mu\nu}$ are
respectively the energy density, pressure, four velocity of the
fluid and bulk and shear viscous stresses. In the LL, the velocity field is
defined in such a way that the energy current vanishes in the local rest
frame, $u^{\mu}\rightarrow\left( 1,0,0,0\right) $. In this local rest frame,
it is assumed that the equation of state and thermodynamical relations are
valid as if the fluid were in equilibrium. As usual, we write 
\begin{equation*}
u^{\mu}=\left( 
\begin{array}{c}
\gamma \\ 
\gamma\vec{v}%
\end{array}
\right),
\end{equation*}
where $\gamma$ is the Lorentz factor and 
\begin{equation*}
u^{\mu}u_{\mu}=1.
\end{equation*}
The tensor $P^{\mu\nu}$ is the projection operator to the space orthogonal
to $u^{\mu}$ and given by 
\begin{equation*}
P^{\mu\nu}=g^{\mu\nu}-u^{\mu}u^{\nu},
\end{equation*}
which has the following trace property, 
\begin{eqnarray}
\sum_{\lambda=0}^{D}P^{\lambda}_{\lambda} = D.
\end{eqnarray}
Here, $D$ is the number of spatial dimensions.

In the LL, the current for the conserved quantity (e.g., net baryon
number) takes the form 
\begin{equation}
N^{\mu}=nu^{\mu}+\nu^{\mu},  \label{NweN}
\end{equation}
where $\nu^{\mu}$ is the heat conduction part of the current. For the
irreversible currents, we require the constraints \cite{LL}, 
\begin{equation}
u_{\mu}\pi^{\mu\nu}=0,  \label{Orthogonal1}
\end{equation}
and 
\begin{equation}
u_{\mu}\nu^{\mu}=0.  \label{Orthogonal2}
\end{equation}
These constraints permit us to interpret $\varepsilon$ and $n$ respectively
as the energy and net baryon number densities in the local rest
frame. In fact, from Eq.(\ref{Orthogonal2}), in the rest frame, we have
\begin{equation*}
N^{\mu}\rightarrow\left( 
\begin{array}{c}
n \\ 
\vec{\nu}%
\end{array}
\right),
\end{equation*}
so that $n$ is the net baryon number density in the local rest
frame. 
It is noteworthy that, different from the ideal case (or
Eckart case), $N^0$ is not related with the net baryon number density $n$ in
any frame by the Lorentz contraction, but 
\begin{equation*}
N^{0}=n\gamma+\left( \vec{\nu}\cdot\vec{v}\right) \gamma.
\end{equation*}

With these irreversible currents, of course, the entropy is not conserved.
Instead, from Eqs.(\ref{divTmunu}) and (\ref{divN}) with the constraints
Eqs. (\ref{Orthogonal1}) and (\ref{Orthogonal2}), we have \cite{LL} 
\begin{equation}
\partial_{\mu}\left( su^{\mu}-\alpha\nu^{\mu}\right) =\frac{1}{T}\left(
-P^{\mu\nu}\Pi+\pi^{\mu\nu}\right)
\partial_{\mu}u_{\nu}-\nu^{\mu}\partial_{\mu}\alpha,  \label{s-current}
\end{equation}
where $\alpha=\mu/T$ with temperature $T$ and chemical potential $\mu$.
Landau-Lifshitz identifies the term 
\begin{equation}
\sigma^{\mu}=su^{\mu}-\alpha\nu^{\mu} ,  \label{sigma}
\end{equation}
as the entropy four-flux. The r.h.s. of Eq. (\ref{s-current}) is the source
term for entropy production.

In non-equilibrium thermodynamics, it is interpreted that entropy production
is the sum of the products of thermodynamic forces and irreversible
currents. Thus, we can define the scalar, vector and tensor thermodynamic
forces, 
\begin{equation*}
F=\partial_{\alpha}u^{\alpha},\ \ \ F_{\mu}=\partial_{\mu}\alpha ,\ \ \
F_{\mu\nu}=\partial_{\mu}u_{\nu},
\end{equation*}
respectively. To satisfy the second law of thermodynamics, we assume that
the entropy production is positive, 
\begin{equation}
\frac{1}{T}\left( -P^{\mu\nu}\Pi+\pi^{\mu\nu}\right) \partial_{\mu}u_{\nu
}-\nu^{\mu}\partial_{\mu}\alpha\geq0.  \label{Landau}
\end{equation}
To maintain this algebraic positive definiteness, the most general
irreversible currents are given by linear combinations of the thermodynamic
forces with the coefficients appropriately chosen. However, if we accept the
Curie (symmetry) principle which forbids the mixture of different types of
thermodynamic forces \cite{Koide3}, the irreversible currents are given by 
\begin{align}
\Pi & =-\zeta F=-\zeta\partial_{\alpha}u^{\alpha},  \notag \\
\pi_{\mu\nu} & =P_{\mu\nu\alpha\beta}\widetilde{\pi}^{\alpha\beta}=\eta
P_{\mu\nu\alpha\beta}F^{\alpha\beta}=\eta P_{\mu\nu\alpha\beta}\partial
^{\alpha}u^{\beta},  \notag \\
\nu_{\mu} & =P_{\mu\nu}\widetilde{\nu}^{\nu}=-\kappa P_{\mu\nu}F^{\nu
}=-\kappa P_{\mu\nu}\partial^{\nu}\alpha,  \label{viscous}
\end{align}
where $\zeta,$ $\eta$ and $\kappa$ are bulk viscosity, shear viscosity and
thermal conductivity coefficients, respectively. Here, $P^{\mu\alpha\nu%
\beta} $ is the double symmetric traceless projection,%
\begin{equation}
P^{\mu\nu\alpha\beta}=\frac{1}{2}\left( P^{\mu\alpha}P^{\nu\beta}+P^{\mu
\beta}P^{\nu\alpha}\right) -\frac{1}{D}P^{\mu\nu }P^{\alpha\beta},
\end{equation}
and we have introduced the quantities $\widetilde{\pi}^{\alpha\beta}$ and $%
\widetilde{\nu}^{\nu}$ which correspond respectively to the shear tensor and
irreversible current before the projection. They are in general not
orthogonal to $u^{\mu}$ so the projection operators are necessary to satisfy
the constraints Eqs.(\ref{Orthogonal1}) and (\ref{Orthogonal2}).

Eqs.(\ref{viscous}) are the prescription of the LL. As mentioned, the LL
leads to the acausal propagation of signal. So we should modify these
equations to satisfy the relativistic causality principle. The basic point
is that the equations of the LL form a parabolic system and we have to
convert it to the hyperbolic one. However, at this moment, the
generalization of these equation in order to obtain hyperbolic equations is
rather self-evident. We introduce the memory function in each irreversible
currents, Eq.(\ref{viscous}), 
\begin{align}
\Pi\left( \tau\right) & =-\int_{-\infty}^{\tau}d\tau^{\prime}G\left(
\tau,\tau^{\prime}\right) \zeta\partial_{\alpha}u^{\alpha}\left(
\tau^{\prime}\right) ,  \notag \\
\widetilde{\pi}^{\mu\nu}\left( \tau\right) & =
\int_{-\infty}^{\tau}d\tau^{\prime}G\left( \tau,\tau^{\prime}\right)
\eta\partial^{\mu}u^{\nu }\left( \tau^{\prime}\right) ,  \notag \\
\widetilde{\nu}^{\mu}\left( \tau\right) &
=-\int_{-\infty}^{\tau}d\tau^{\prime}G\left( \tau,\tau^{\prime}\right)
\kappa\partial^{\mu}\alpha\left( \tau^{\prime}\right) ,  \label{Integrals}
\end{align}
where $\tau=\tau\left( \vec{r},t\right) $ is the local proper time. As
before, the shear tensor $\pi^{\mu\nu}$ and the irreversible current $\nu
^{\mu}$ are then given by the projection of these integrals as%
\begin{align}
\pi^{\mu\nu} & =P^{\mu\nu\alpha\beta}\widetilde{\pi}_{\alpha\beta}\left(
\tau\right) ,  \notag \\
\nu^{\mu} & =P^{\mu\nu}\widetilde{\nu}_{\nu}\left( \tau\right) .
\label{Projection}
\end{align}

When we start with the finite initial time, say, $\tau_{0},$ the above
integrals should read%
\begin{align}
\Pi\left( \tau\right) & =-\int_{\tau_{0}}^{\tau}d\tau^{\prime}G\left(
\tau,\tau^{\prime}\right) \zeta\partial_{\alpha}u^{\alpha}\left(
\tau^{\prime}\right) +e^{-(\tau-\tau_{0})/\tau_{R}}\Pi_{0}, \\
\widetilde{\pi}^{\mu\nu}\left( \tau\right) & =
\int_{\tau_{0}}^{\tau}d\tau^{\prime}G\left( \tau,\tau^{\prime}\right)
\eta\partial^{\mu}u^{\nu }\left( \tau^{\prime}\right)
+e^{-(\tau-\tau_{0})/\tau_{R}}\widetilde{\pi}^{\mu\nu}{}_{0}, \\
\widetilde{\nu}^{\mu}\left( \tau\right) &
=-\int_{\tau_{0}}^{\tau}d\tau^{\prime}G\left( \tau,\tau^{\prime}\right)
\kappa\partial^{\mu}\alpha\left( \tau^{\prime}\right)
+e^{-(\tau-\tau_{0})/\tau_{R}}\widetilde{\nu}^{\mu}{}_{0},
\end{align}
where $\Pi_{0},$ $\widetilde{\pi}^{\mu\nu}{}_{0}$ and $\widetilde{\nu}%
^{\mu}{}_{0}$ are initial conditions given at $\tau_0$. 
The bulk
viscosity can be determined by $T^{\mu\nu}\left( \tau_{0}\right) $ and $
N^{\mu }\left( \tau_{0}\right) $ as%
\begin{align*}
\Pi_{0} & = \frac{1}{D}(\varepsilon_0 - T^{\mu}_{\mu}(\tau_0)) - p_0.
\end{align*}
As for $\widetilde{\pi}^{\mu\nu}{}_{0}$ and $\widetilde{\nu}^{\mu}{}_{0}$ we
observe that they are the projected part of $T^{\mu\nu}\left(
\tau_{0}\right) $ and $N^{\mu }\left( \tau_{0}\right)$, respectively, 
\begin{align*}
{\nu}^{\mu}{}_{0} & =N^{\mu}\left( \tau_{0}\right)
-n_{0}u_{0}^{\mu}=P^\mu_\nu N^\nu(\tau_0), \\
{\pi}^{\mu\nu}{}_{0} & =T^{\mu\nu}\left( \tau_{0}\right) -\left(
\varepsilon_{0}+p_{0}+\Pi_{0}\right) u_{0}^{\mu}u_{0}^{\nu}+(p_{0}+\Pi
_{0})g^{\mu\nu}=P^{\mu\nu}_{\alpha\beta}T^{\alpha\beta}(\tau_0).
\end{align*}
This is equivalent to set up the following initial conditions, 
\begin{align*}
\widetilde{\nu}^{\mu}{}_{0} & =N^{\mu}\left( \tau_{0}\right)
-n_{0}u_{0}^{\mu}, \\
\widetilde{\pi}^{\mu\nu}{}_{0} & =T^{\mu\nu}\left( \tau_{0}\right) -\left(
\varepsilon_{0}+p_{0}+\Pi_{0}\right) u_{0}^{\mu}u_{0}^{\nu}+(p_{0}+\Pi
_{0})g^{\mu\nu}.
\end{align*}
The initial four-velocity $u_{0}^{\mu}$ is determined as the time-like
eigenvector of the matrix $T^{\mu\nu}\left( \tau_{0}\right) $ with
eigenvalue $\varepsilon_{0}.$ Using the equation of state together with $%
n_{0}=u_{\mu}N^{\mu},$ we can determine the pressure $p_{0.}$ 

In Eqs.(\ref{Integrals}), we have used the same memory function $G$ and
consequently a common relaxation time $\tau_{R}$ for the bulk and shear
viscosities and heat conduction. We could have used different relaxation
times for each irreversible current and this would not alter the basic
structure of our theory. However, here we stay with a common relaxation time
for all of them for the sake of simplicity. We consider the situation where
the time scales of the microscopic degrees of freedom are well separated
from those of the macroscopic ones. Then, the effect of the differences of
the microscopic relaxation times should not be much relevant in the dynamics
described in the macroscopic time scale. Thus we just represent these
microscopic time scales in terms of one relaxation time $\tau_{R}.$

The integral expressions (\ref{Integrals}) are equivalent to the following
differential equations, 
\begin{align}
\Pi & =-\zeta \partial _{\alpha }u^{\alpha }-\tau _{R}\frac{d\Pi }{d\tau }, 
\notag \\
\widetilde{\pi }^{\mu \nu }& =\eta \partial ^{\mu }u^{\nu }-\tau _{R}\frac{d%
\widetilde{\pi }^{\mu \nu }}{d\tau },  \notag \\
\widetilde{\nu }^{\mu }& =-\kappa \partial ^{\mu }\alpha -\tau _{R}\frac{d%
\widetilde{\nu }^{\mu }}{d\tau },  \label{derivatives}
\end{align}%
where 
\begin{equation*}
\frac{d}{d\tau }=u^{\mu }\partial _{\mu },
\end{equation*}%
is the total derivative with respect to the proper time. In a
practical implementation of our theory, we may solve the above differential
equations together with the other hydrodynamical part. In this case, if we
wish, we may add also the terms which violates the Curie principle, without
any extra difficulties. 
The above equations, after the projection (\ref{Projection}), 
can be compared to the corresponding equations in the
simplest version of the IS, which are obtained phenomenologically based on
extended thermodynamics, 
\begin{align}
\Pi _{IS}& =-\frac{1}{3}\zeta _{IS}\left( \partial _{\alpha }u^{\alpha
}+\beta _{0}\frac{d\Pi _{IS}}{d\tau }-\alpha _{0}\partial _{\alpha }\nu
^{\alpha }\right) ,  \notag \\
\pi _{IS}^{\mu \nu }& =2\eta _{IS}P^{\mu \alpha \nu \beta }\left( \partial
_{\alpha }u_{\beta }-\beta _{2}\frac{d\pi _{IS}^{\mu \nu }}{d\tau }-\alpha
_{1}\partial _{\alpha }\left( \nu _{IS}\right) _{\beta }\right) ,  \notag \\
\nu _{IS}^{\mu }& =-\kappa _{IS}P^{\mu \nu }\left( \frac{n}{\varepsilon +P}%
\partial _{\nu }\alpha +\beta _{1}\frac{d\nu _{IS}^{\nu }}{d\tau }+\alpha
_{0}\partial _{\nu }\Pi _{IS}+\alpha _{1}\partial _{\alpha }\left( \pi
_{IS}\right) _{\nu }^{\alpha }\right) ,  \label{ISM}
\end{align}%
where $\alpha _{0},\alpha _{1},\beta _{0},\beta _{1}$ and $\beta _{2}$ are
constants. Note that the definitions of parameters $\eta $, $\zeta $ and $%
\kappa $ are different from that of the IS. Eqs.(\ref{derivatives}) and (\ref%
{ISM}) have similar aspects, in particular, if we include the terms
which violate the Curie principle and introduce the 3 different relaxation
times. 
However, our equation is \textit{not} a special case of the
IS. In the IS, the projection operators, which are necessary to satisfy the
orthogonality conditions (\ref{Orthogonal1}) and (\ref{Orthogonal2}), are
included in the differential equations themselves. Thus, it is not possible
to derive our equations from the IS. For example, in our theory, we can
write down the differential equation of the heat conduction $\nu ^{\mu }$ by
using Eq. (\ref{derivatives}) as follows, 
\begin{equation*}
\nu ^{\mu }=-\kappa P^{\mu \nu }\partial _{\nu } \alpha -\tau _{R}
\frac{d\nu^{\mu }}{d\tau } + \frac{dP^{\mu \nu }}{d\tau }\widetilde{\nu }_{\nu }.
\end{equation*}
The last term of the above equation do not appear in the IS.

In practice, the differential equations for the irreversible
currents can be solved together with the equations of motion of the usual
hydrodynamic variables in a coupled way. In the case of the IS, due to the
presence of projection operators, these irreversible current should be
determined simultaneously with the equations for acceleration of the fluid.
Thus, in the general 3D case, $\left( 14\times 14\right) $ matrix
inversion is required in each time step.  In our case, due to the memory
effect integral, the equation for the acceleration of fluid can be
determined only from the past values of the irreversible currents, so that
we need the inversion of at most $\left( 3\times 3\right) $ matrix
for the acceleration of the fluid. The time derivatives for the irreversible
currents are decoupled from the normal hydrodynamic variables.

In spite of these differences, our equations are found to be still
hyperbolic in the linearized form. When we consider the propagation of small
perturbations on the homogeneous and static background, the projection
operator turns out to be a constant matrix. Therefore we can easily see that
our linearized equation of motion has the same structure as the IS with $%
\alpha_{0}=\alpha_{1}=0$. Thus the speed of pulse propagation is finite as
discussed by Hiscock-Lindblom \cite{Hiscock1,Hiscock2,Hiscock3}.

In our case, we can explicit the space component of the
four-divergence of the shear tensor $\pi^{\mu \nu }$ can be written as 
\begin{align}
\partial _{\mu }\pi ^{\mu i}& =-\sigma \frac{d}{d\tau }\frac{1}{\sigma }%
\left[ u_{\alpha }\widetilde{\pi }^{\left( \alpha \ i\right) }+\left\{ \frac{%
1}{\gamma }u_{\alpha }\widetilde{\pi }^{\left( \alpha \ 0\right) }-\frac{2}{3%
}\left( u_{\alpha }u_{\beta }\widetilde{\pi }^{\left( \alpha \ \beta \right)
}\right) \right\} u^{i}\right]  \notag \\
& +\frac{d\pi ^{\left( \mu \ i\right) }(\tau )}{d\tau }\partial _{\mu }\tau
-\nabla \cdot \left( u^{i}u_{\alpha }\left( \vec{\pi}^{\alpha }-\widetilde{%
\pi }^{\left( \alpha \ 0\right) }\vec{v}\right) \right) +\frac{1}{3}\partial
^{i}(P_{\alpha \beta }\widetilde{\pi }^{(\alpha \beta )})+\frac{1}{3}\sigma 
\frac{d}{d\tau }\frac{u^{i}\widetilde{\pi }_{\alpha }^{\alpha }}{\sigma },
\label{dtaumunu}
\end{align}%
where we have introduced the symmetric tensor, 
\begin{equation*}
\widetilde{\pi }^{\left( \alpha \ \beta \right) }=\frac{1}{2}\left( 
\widetilde{\pi }^{\alpha \beta }+\widetilde{\pi }^{\beta \alpha }\right),
\end{equation*}%
and used the three-vector notation,%
\begin{equation*}
\vec{\pi}^{\alpha }=\left( 
\begin{array}{c}
\widetilde{\pi }^{\left( \alpha \ 1\right) } \\ 
\widetilde{\pi }^{\left( \alpha \ 2\right) } \\ 
\widetilde{\pi }^{\left( \alpha \ 3\right) }%
\end{array}%
\right) .
\end{equation*}%
Furthermore, for the sake of later convenience we have introduced the
\textquotedblleft reference density\textquotedblright\ $\sigma =\sigma
\left( t,\vec{r}\right) $ defined by%
\begin{equation}
\frac{1}{\sigma }\frac{d\sigma }{d\tau }=-\partial _{\mu }u^{\mu }.
\label{refdens}
\end{equation}%
This will play an important role for the smoothed particle hydrodynamics
(SPH) formulation \cite{Spherio,Review}. Note that, in general for any
function $f=f\left( t,\vec{r}\right) $ we have the identity, 
\begin{equation}
\partial _{\mu }\left( fu^{\mu }\right) =\sigma \frac{d}{d\tau }\left( \frac{%
f}{\sigma }\right) .  \label{div}
\end{equation}

The equation of motion finally has the form, 
\begin{align}
& \frac{d}{d\tau}\frac{1}{\sigma}\left[ \left\{ \varepsilon+p+\Pi-\frac {1}{%
\gamma}u_{\alpha}\widetilde{\pi}^{\left( \alpha\ 0\right) } + \frac{1}{3}%
\widetilde{\pi}^{\alpha}_{\alpha} +\frac{2}{3}\left( u_{\alpha}u_{\beta}%
\widetilde{\pi}^{\left( \alpha\ \beta\right) }\right) \right\}
u^{i}-u_{\alpha}\widetilde{\pi}^{\left( \alpha\ i\right) }\right]  \notag \\
& =-\frac{1}{\sigma}\partial_{i}\left( p+\Pi+\frac{1}{3}\left(
P^{\alpha\beta}\widetilde{\pi}_{\left( \alpha\ \beta\right) }\right) \right)
-\frac{1}{\sigma}\frac{d\widetilde{\pi}^{\left( \mu\ i\right) }(\tau)}{d\tau 
}\partial_{\mu}\tau +\frac{1}{\sigma}\nabla\cdot\left( u^{i}u_{\alpha}\left( 
\vec{\pi }^{\alpha}-\widetilde{\pi}^{\left( \alpha\ 0\right) }\vec{v}\right)
\right).  \label{EqMotion}
\end{align}
Note that the right-hand side of the above equation does not contain the
acceleration. From this equation, we can determine the time derivative of $%
u^{i},$ $i=1,2,3.$

The time component is equivalent to the energy (entropy) equation, 
\begin{equation}
\partial_{\mu}(su^{\mu})=\frac{1}{T}\left( -P^{\mu\nu}\Pi+\pi^{\mu\nu
}\right) \partial_{\mu}u_{\nu}+\alpha\partial_{\mu}\nu^{\mu}.  \label{dmusmu}
\end{equation}
so that using Eq.(\ref{div}), we have%
\begin{equation}
\frac{d}{d\tau}\left( \frac{s}{\sigma}\right) =\frac{1}{\sigma T}\left(
-P^{\mu\nu}\Pi+\pi^{\mu\nu}\right) \partial_{\mu}u_{\nu}+\alpha\partial_{\mu
}\nu^{\mu}.  \label{Entropy}
\end{equation}
The conservation of net baryon number is written as%
\begin{equation}
\frac{d}{d\tau}\left( \frac{n}{\sigma}\right) =-\frac{1}{\sigma}%
\partial_{\alpha}\nu^{\alpha}.  \label{baryon}
\end{equation}
Eqs.(\ref{refdens},\ref{EqMotion},\ref{Entropy},\ref{baryon}) specify the
time evolution of $u^{i},s,n$ and $\sigma$. These equations constitute a
closed system with the use of the equation of state and thermodynamic
relations, $\mu=\partial\varepsilon/\partial n$ and $T=\partial\varepsilon/%
\partial s,$ and also the integral expressions, Eq.(\ref{Integrals}) for $%
\Pi,$ $\widetilde{\pi}^{\mu\nu}$ and $\widetilde{\nu}^{\mu}$ (and their
projected tensors, $\pi^{\mu\nu}$ and $\nu^{\mu}$).

From the practical point of view, our approach has several advantages to the
IS in addition to its simple physical structure. Although the IS invokes the
information from the kinetic theory, it is not probable that we can
determine the parameters from the relativistic heavy ion processes
quantitatively at the present moment. From the theoretical point of view,
one could try to fix these parameters by using the Boltzmann equation
approach to QCD. However, the Boltzmann equation is applicable to describe
the behavior of a gas and if the QGP is a strongly interacting fluid, then
such an approach will fail. Rather, we prefer to keep the simplest physical
structure of the Navier-Stokes theory preserving causality. To accomplish
this, we simply introduced the memory effect characterized by the relaxation
time. All irreversible currents are then expressed by Eq.(\ref{Integrals}).
We only need the past values of the independent variables of the usual ideal
fluid dynamics to advance in time. The integral expressions (\ref{Integrals}%
) are easy to be evaluated when we use the Lagrangian coordinate system.
Because of the simple form of the dissipative terms, it is straightforward
to incorporate these equations to the realistic full three-dimensional
hydro-code such as SPheRIO \cite{Spherio,Review}.

Until now, we have considered that the relaxation time $\tau_{R}$ is
constant. However in practical problems, it is a function of thermodynamical
variables. Then the memory function should be generalized as%
\begin{equation}
G\left( \tau,\tau^{\prime}\right) \rightarrow\frac{1}{\tau_{R}\left(
\tau^{\prime}\right) }e^{-\int_{\tau^{\prime}}^{\tau}\frac{1}{\tau_{R}\left(
\tau^{\prime\prime}\right) }d\tau^{\prime\prime}}.
\end{equation}
Even in this case, Eqs.(\ref{Integrals}) and (\ref{EqMotion}) are still
valid.

\section{Entropy Production}

It should be emphasized that our theory is not a simplified version of the
IS but there exists an essential difference for the treatment of the entropy
production term. The IS requires the general algebraic form of the
non-negative definite expression for entropy production following
non-equilibrium thermodynamics. In our approach, we have relaxed this
condition, that is, the expression Eq.(\ref{Landau}) for the entropy
production 
\begin{equation*}
\frac{1}{T}\left( -P^{\mu \nu }\Pi +\pi ^{\mu \nu }\right) \partial _{\mu
}u_{\nu }-\nu ^{\mu }\partial _{\mu }\alpha \geq 0,
\end{equation*}%
does not guarantee \textit{algebraically} owing to the non-locality in time
contained in $\Pi ,$ $\pi ^{\mu \nu }$ and $\nu ^{\mu }$ through Eq.(\ref%
{Integrals}). This might seem to be dangerous. However, strictly speaking,
the increase of entropy is essentially a concept in the equilibrium
thermodynamics and the requirement of positiveness should apply only to
thermal equilibrium states. As a matter of fact, it was recently shown that
the entropy absorption process can occur in the non-equilibrium processes of
mesoscopic systems \cite{fluctuation}. In our approach, we are dealing with
a fluid element which is out of equilibrium, interacting with the
neighboring elements. Therefore, within the relaxation time, its entropy
content may increase or decrease depending on the dynamics and its time
scales. Thus, the requirement of the algebraic positive definiteness
irrespective of any field configuration seems to be too restrictive. The
requirement of non-negative entropy production may be relaxed for far from
equilibrium states. In our case, apart from the projection operators, the
expression for entropy production has the form 
\begin{equation}
Q\left( \tau \right) =f\left( \tau \right) \frac{1}{\tau _{R}}\int^{\tau
}d\tau ^{\prime }e^{-\left( \tau -\tau ^{\prime }\right) /\tau _{R}}f\left(
\tau ^{\prime }\right) ,
\end{equation}%
where $f$ is one of $\partial _{\mu }u_{\nu }$ or $\partial _{\mu }\alpha .$
For small $\tau _{R}$, we may expand $f\left( \tau ^{\prime }\right) $ near $%
\tau ,$ 
\begin{equation}
f\left( \tau ^{\prime }\right) =f\left( \tau \right) -\left( \tau -\tau
^{\prime }\right) \frac{df\left( \tau \right) }{d\tau }+\cdots ,
\end{equation}%
and we have 
\begin{equation}
Q\left( \tau \right) =f\left( \tau \right) \left[ f\left( \tau \right) -\tau
_{R}\frac{df\left( \tau \right) }{d\tau }+O\left( \tau _{R}^{\ 2}\right) %
\right] .
\end{equation}%
Thus, as far as 
\begin{equation}
\left\vert \tau _{R}\frac{df\left( \tau \right) }{d\tau }\right\vert
<\left\vert f\left( \tau \right) \right\vert ,
\end{equation}%
the positiveness of the entropy is ensured. The l.h.s. of the above equation
is the amount of variation of $f\left( \tau \right) $ within a small time
interval $\tau _{R}$. Thus, the above condition shows that if the time
evolution of the system is not too violent (the change of field values
within the relaxation time is less than its value), then the local entropy
production is not negative. For the example discussed below, we can show
explicitly the positive definiteness of entropy production on our
formulation.

In the above, we considered the relaxation time as constant just for the
illustration. The similar conclusion can be derived when the variation of
the relaxation time is not so violent.

\section{One-Dimensional Scaling Solution}

To see how the above scheme works, let us apply it to the one dimensional
scaling solution of the Bjorken model. This has been studied already in the
framework of the IS \cite{Muronga1,Muronga2,Muronga3,Baier1,Baier2}. The
components of the irreversible currents is written down explicitly as 
\begin{align}
\Pi\left( \tau\right) & =-\int_{\tau_{0}}^{\tau}d\tau^{\prime}G\left(
\tau,\tau^{\prime}\right) \frac{\zeta}{\tau^{\prime}}+\tau_{R}\left(
\tau_{0}\right) G\left( \tau,\tau_{0}\right) \Pi\left( \tau_{0}\right) , \\
\Omega\left( \tau\right) & =-\int_{\tau_{0}}^{\tau}d\tau^{\prime}G\left(
\tau,\tau^{\prime}\right) \frac{\eta}{\tau^{\prime}}+\tau_{R}\left( \tau
_{0}\right) G\left( \tau,\tau_{0}\right) \Omega\left( \tau_{0}\right) , \\
\Phi\left( \tau\right) & =-\int_{\tau_{0}}^{\tau}d\tau^{\prime}G\left(
\tau,\tau^{\prime}\right) \kappa\frac{d\alpha}{d\tau^{\prime}}+\tau
_{R}\left( \tau_{0}\right) G\left( \tau,\tau_{0}\right) \Phi\left(
\tau_{0}\right) , \\
\left( \widetilde{\pi}_{\mu\nu}\right) & =-\left( 
\begin{array}{cc}
-\sinh^{2}y & \sinh y\cosh y \\ 
\sinh y\cosh y & -\cosh^{2}y%
\end{array}
\right) \Omega\left( \tau\right) , \\
\left( \widetilde{\nu}_{\mu}\right) & =\left( 
\begin{array}{c}
\cosh y \\ 
-\sinh y%
\end{array}
\right) \Phi\left( \tau\right),
\end{align}
where we have used the hyperbolic variables,%
\begin{equation*}
t=\tau\cosh y,\ x=\tau\sinh y,
\end{equation*}
and used the scaling ansatz in $y$ (that is, there is no $y$ dependence in
thermodynamic variables). $\Pi\left( \tau_{0}\right) ,\Omega\left( \tau
_{0}\right) \ $and$\ \Phi\left( \tau_{0}\right) $ are initial values for $%
\Pi\left( \tau\right) ,\Omega\left( \tau\right) \ $and$\ \Phi\left(
\tau\right) .$ We obtain 
\begin{align}
\pi^{\mu\nu} & =P^{\mu\alpha\nu\beta}\widetilde{\pi}_{\alpha\beta}=-\frac{%
2\Omega}{3}P^{\mu\nu}, \\
\nu^{\mu} & =P^{\mu\nu}\widetilde{\nu}_{\mu}=0.
\end{align}

As we see, in the one-dimensional case, if $\zeta$ and $\eta$ are
proportional as functions of thermodynamic quantities such as $T$ and $\mu,$
then the bulk and shear viscosity terms are not independent, and 
\begin{equation*}
\Pi\propto\Omega.
\end{equation*}
However, when $\zeta$ and $\eta$ have, in general, different dependence on
the thermodynamic quantities, the two viscosities act differently.

The time component of the divergence of $T^{\mu\nu}$ gives 
\begin{equation}
\frac{d}{d\tau}\epsilon\left( \tau\right) +\frac{\epsilon+p+\Pi}{\tau}+\frac{%
2}{3}\frac{\Omega}{\tau}=0.  \label{dy}
\end{equation}
The equation for the space component is automatically satisfied by the
scaling ansatz showing its consistency. The entropy production rate is
calculated to be 
\begin{equation}
\partial_{\mu}(su^{\mu}-\alpha\nu^{\mu})=-\frac{1}{T}\frac{1}{\tau}\left(
\Pi+\frac{2}{3}\Omega\right) .
\end{equation}
Since $\Pi$ and $\Omega$ are negative definite, the entropy production is
positive definite.

\subsection{Solutions}

When $\zeta =\zeta _{0},$ $\eta =\eta _{0}$ and $\tau _{R}\ $are constant,
then we can obtain analytic expression for the proper energy density. We
obtain 
\begin{equation}
\Omega =\frac{\eta _{0}}{\zeta _{0}}\Pi =-\frac{\eta _{0}}{\tau _{R}}e^{-%
\frac{\tau }{\tau _{R}}}\left[ \mathrm{Ei}\left( -\frac{\tau }{\tau _{R}}%
\right) -\mathrm{Ei}\left( -\frac{\tau _{0}}{\tau _{R}}\right) +E_{0}\right]
,
\end{equation}%
where 
\begin{equation*}
\mathrm{Ei}\left( -x\right) =\int_{x}^{\infty }\frac{e^{-t}}{t}dt,
\end{equation*}%
is the exponential integral, and $E_{0}$ is a constant which should be
determined from the initial condition for $\Pi $ (or $\Omega $). For
relativistic ideal gas, 
\begin{equation*}
P=\frac{\varepsilon }{3},
\end{equation*}%
we get 
\begin{equation}
\frac{d\varepsilon }{d\tau }+\frac{4}{3}\frac{\varepsilon }{\tau }+\left( 
\frac{2\eta _{0}}{3\zeta _{0}}+1\right) \frac{\Pi \left( \tau \right) }{\tau 
}=0,
\end{equation}%
so that for $E_{0}=0,$ 
\begin{equation}
\varepsilon \left( \tau \right) =\varepsilon _{_{0}}\left( \frac{\tau _{_{0}}%
}{\tau }\right) ^{4/3}\left[ 1-\frac{1+2\eta _{0}/3\zeta _{0}}{\varepsilon
_{_{0}}\tau _{_{0}}^{4/3}}\int_{\tau _{0}}^{\tau }dtt^{1/3}\Pi \left(
t\right) \right] ,  \label{sol1}
\end{equation}%
where the integral containing the exponential function can still be
evaluated analytically. The temperature is determined from the energy
density as 
\begin{equation*}
\varepsilon =\sigma _{SB}T^{4},
\end{equation*}%
where $\sigma _{SB}$ is the Stephan-Boltzmann constant.

On the other hand, a typical estimate from the kinetic theory shows that the
shear viscosity $\eta$ is proportional to the entropy density $s$,$\ \eta
=bs $, where $b$ is a constant \cite{Muronga2,Baier1}. Following Ref. \cite%
{Baier1}, we choose $b=1.1$. Furthermore, we use the relaxation time \cite%
{Muronga2,Baier1} 
\begin{equation}
\tau_{R}=\frac{3\eta_{IS}}{4p}=\frac{3\eta}{8p}.  \label{RT}
\end{equation}
Here, it should be noted that our definition of $\eta$ is twice of other
papers \cite{Muronga1,Muronga2,Muronga3,Baier1,Baier2}. The effect of the
bulk viscosity has not been discussed in previous papers. We analogously
assume that the bulk viscosity has the similar $s$ dependence, $\zeta=as$.
For a baryon free relativistic gas, $s$ is related as the energy density as%
\begin{equation*}
s=C\varepsilon^{3/4},
\end{equation*}
so that the equation for the energy density becomes%
\begin{equation}
\tau\frac{d^{2}\varepsilon}{d\tau^{2}}+\left( \frac{7}{3}+\frac{\tau}{
\tau_{R}}\right) \frac{d\varepsilon}{d\tau}+\frac{1}{\tau_{R}}\left( \frac{4 
}{3}\varepsilon-C^{\prime}\frac{\varepsilon^{3/4}}{\tau}\right) =0,
\label{sol2}
\end{equation}
were 
\begin{equation*}
C^{\prime}=\left( a+\frac{2}{3}b\right) C.
\end{equation*}
The above equation is the same as the equation derived in Ref. \cite{Baier1}%
. It should be noted that this coincidence is due to the specific property
of this particular model. In Ref. \cite{Baier1}, the above equation is
obtained under the assumption of no acceleration condition, 
\begin{equation*}
\frac{d}{d\tau}u^{\mu}=0,
\end{equation*}
which is automatically satisfied in the scaling solution. 
It should
be noted that in the IS, this condition is not satisfied in general, and one
has to solve the original equation of the IS without using the above
condition. However, our theory does not require such a condition at all to
be applied. 
Eq.(\ref{sol2}) can be solved for the initial condition, 
\begin{equation*}
\varepsilon=\varepsilon\left( \tau_{0}\right),
\end{equation*}
and%
\begin{equation}
\left. \frac{d\varepsilon}{d\tau}\right\vert _{\tau=\tau_{0}}=-\frac{1}{%
\tau_{0}}\left[ \frac{4}{3}\varepsilon_{0}+\Pi\left( \tau_{0}\right) +\frac{2%
}{3}\Omega\left( \tau_{0}\right) \right] .  \label{iniener}
\end{equation}

Now, we show our numerical results. To compare the previous works, we ignore
the bulk viscosity. In Fig. 1, we show the energy density $\varepsilon $
obtained by solving Eq.(\ref{sol2}) as function of proper time $\tau$. As
the initial condition, we set $\varepsilon (\tau_0) = 1~\mathrm{GeV/fm}^3$, $%
\Pi \left( \tau_{0}\right) =\Omega \left( \tau _{0}\right) =0$ at the
initial proper time $\tau_0 = 0.1$ fm/c. The first two lines from the top
represents the results of the LL. The next two lines shows the results of
our theory. The last line is the result of ideal hydrodynamics. For the
solid lines, we calculated with the viscosity and relaxation time which
depend on temperature (\ref{iniener}). Initially, the effect of viscosity is
small because of the memory effect, the behavior of our theory is similar to
that of ideal hydrodynamics. After the time larger than the relaxation time,
the memory effect is not effective anymore and the behavior is similar to
the result of the LL. As we have mentioned, the behavior of our theory is
the same as the result obtained in Ref. \cite{Baier1} in this case. For the
dashed lines, we calculated with the constant viscosity and relaxation time, 
$\eta=\eta(\varepsilon_0)$ and $\tau_R = \tau_R(\varepsilon_0)$. In this
case, the viscosity is constant so that the heat production stays longer and
has a smaller slope as function of time asymptotically.

\begin{figure}[tbp]
\includegraphics[scale=0.6]{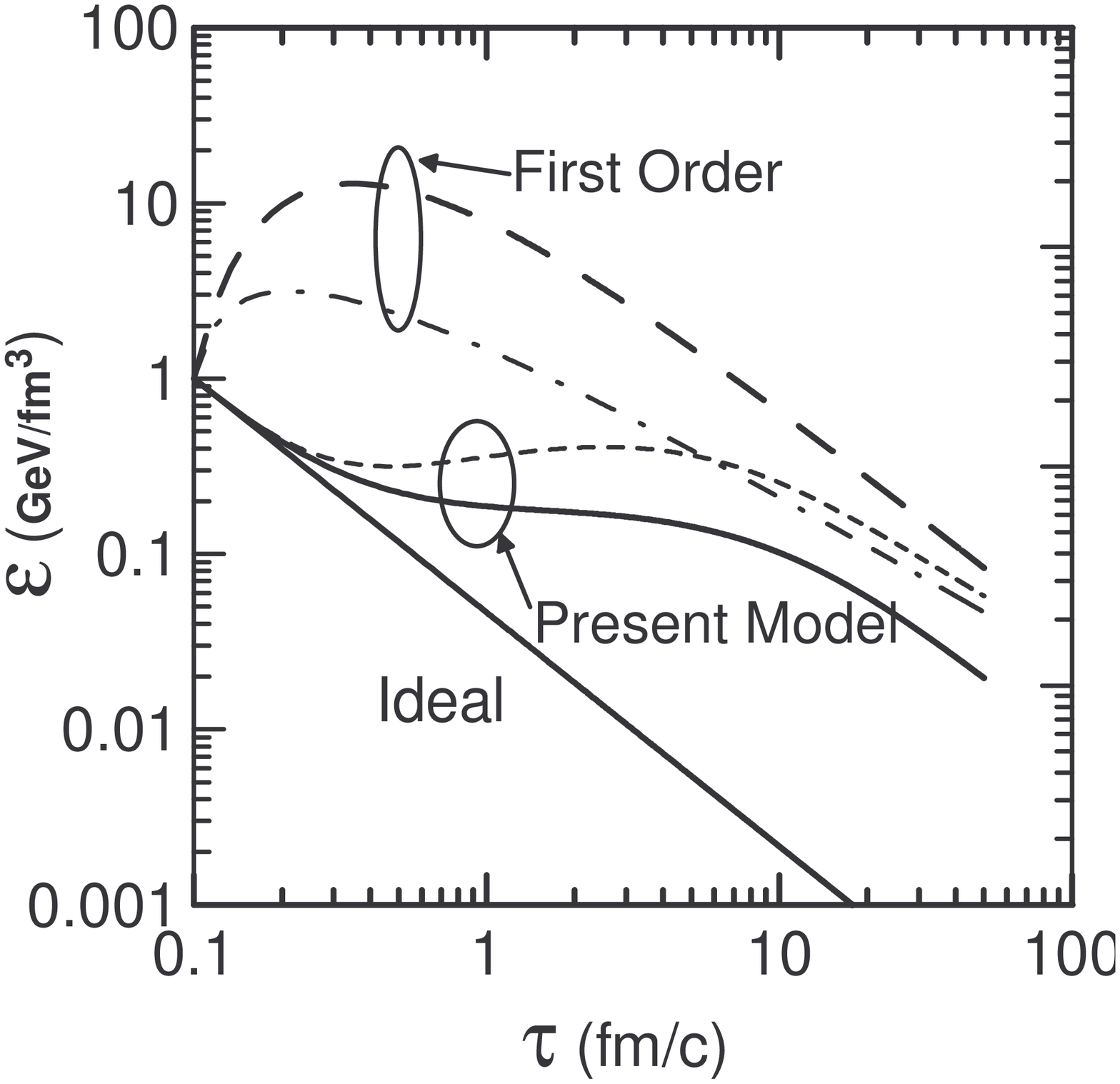}
\caption{The time evolution of the energy density. The dashed curves
correspond to the calculations with the constant viscosity and relaxation
time. The first two lines from the top represent the results of the LL. Next
two lines shows the results of our theory. The last line is the result of
ideal hydrodynamics. }
\label{FIG1}
\end{figure}

Sometimes the emergence of the initial heat-up in the LL (the dashed curve
in Fig.1) is interpreted as an intrinsic problem of the first order theory.
However, such behavior can also appear even in the second order theory. In
Fig. 2, we set $\Pi \left( \tau_{0}\right) = \zeta(\tau_0)/\tau_0$ and $%
\Omega \left( \tau _{0}\right) = \eta(\tau_0)/\tau_0$ as the initial
conditions. In particular, the initial heat-up also appear in the second
order depending on the initial condition for the irreversible currents (see
Fig. 2). Therefore, this heat-up is not the problem of the first order
theory but rather the specific property of the scaling ansatz. This was
already pointed out by Muronga. The physical reason for this heat-up is due
to the use of the Bjorken solution for the velocity field. In this case, the
system acts as if an external force is applied to keep the velocity field as
a given function of $\tau .$ Thus, depending on the relative intensity of
the viscous terms compared to the pressure, the external work converted to
the local heat production can overcome the temperature decrease due to the
expansion.

\begin{figure}[tbp]
\includegraphics[scale=0.6]{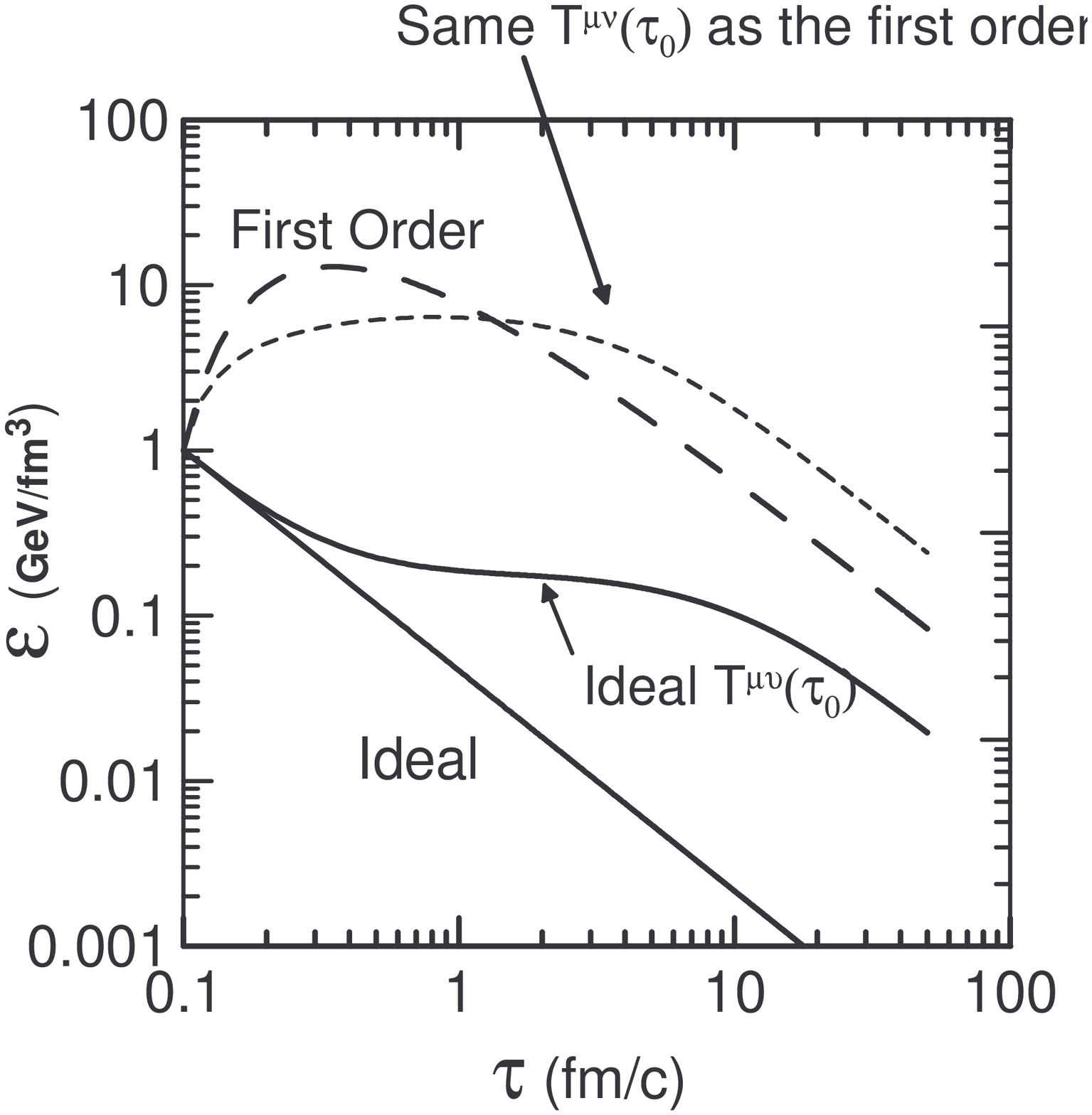}
\caption{ The time evolution of energy density with the different initial
conditions from Fig. 1. The dashed and short dashed lines represent the
result of the LL and our theory, respectively . For comparison, our result
of Fig. 1 is shown, again (ideal $T^{\protect\mu\protect\nu}(\protect\tau_0)$%
). The last line from the top is the result of ideal hydrodynamics. In this
case, the energy heat-up is observed even in our theory. }
\label{FIG2}
\end{figure}

\section{Summary and Concluding remarks}

One of the most important questions in the relativistic heavy ion physics is
to determine the effect of dissipation in flow dynamics of plasma of quarks
and gluons. However, a consistent calculation of relativistic dissipative
hydrodynamics is not trivial at all. Some important questions such as
propagation of shock wave in the hot QCD matter require a careful treatment
of the dissipative processes. The Israel-Stewart theory and the divergence
type theory contain many parameters difficult to be determined from QCD
point of view and also it is technically difficult to be implemented on
practical calculations in its full form \cite{Chau, Heinz, Baier1, Baier2}.

From the practical point of view, it is not desirable to deal with many of
such parameters, since the hydrodynamical approach to the relativistic heavy
ion reaction processes itself has already many uncertainties \cite{OpenProb}%
. In this sense, the IS is not easy to apply at the present stage of the
analysis of heavy ion physics data.

Furthermore, when we deal with numerical solutions, we still encounter
ambiguities. Because dissipations and relaxation times, required to
reproduce the bulk properties of fluid, are not necessarily the same as
those determined from the microscopic theory such as Lattice QCD
calculations. In a practical numerical calculation, the truncation of
hydrodynamic degrees of freedom to the discredited variables introduces a
natural cut-off frequencies, and the elimination of those degrees of freedom
such as small scale vorticities or turbulences will appear as effective
viscosities \cite{turbulence}. Furthermore, the standard numerical method
for dealing with shock phenomena is to introduce the pseudo-viscosity, first
invented by von Neumann and Richtmyer\cite{Neumann}. It is basically the
bulk viscosity depending on the size of the hydro cell. For relativistic
shock motion, we need to incorporate such mechanism to the hydrodynamical
code. Considering these aspects, it is desirable to study the effect of
dissipation with the minimum number of parameters which characterize the
physical processes involved and is still consistent with the framework of
relativity.

In this paper, we proposed an alternative approach to this question,
different from the IS. We start from the physical analysis of the
irreversible currents according to the Landau-Lifshitz theory. Then, the
irreversible currents are given by integral expressions which take into
account the relaxation time. In this way, causality is recovered and at the
same time a simple physical structure of the LL is preserved. In our
approach, only one additional parameter was introduced, the relaxation time, 
$\tau_{R}$. The resulting equation of motion then becomes hyperbolic and
causality can be restored \cite{MullerReviw}. Naturally, causality depends
on the choice of the values of the parameters including the relaxation time.

More specifically, we verified that the linearized equation of motion for
small perturbations in the homogeneous, static background coincides with
Hiscock-Lindblom \cite{Hiscock1,Hiscock2,Hiscock3} except for the coupling
among the different irreversible currents. These couplings are not included
in our theory considering the Curie principle. Of course the Curie principle
is believed to be valid in the regime of the first order theory and in the
second order regime these couplings might be present. However, the existence
of the Curie principle may imply that these couplings are small compared
with the direct terms.

In addition to its simplicity compared to the IS, our formalism has several
practical advantages. First, the number of independent variables is kept the
same as those of the ideal hydrodynamics. The irreversible currents are
expressed explicitly as memory integrals of these independent variables. In
extended (irreversible) thermodynamics, the currents are treated as extended
thermodynamic variables and thermodynamic relations are modified \cite%
{Jou,EOSJOU}. Thus, in the IS, we should take the modification of
thermodynamic relations into account but in our case the usual
thermodynamical relations should be used. Second, in the IS, the projection
operators enters in the equations of motion in a complicated way so it is
not trivial to extract the standard form to apply the conventional method of
the time integration. In our case, this does not occur. Instead, we need
just some additional integrals associated with the local comoving coordinates.
%Third, this numerical integration procedure works even in the vanishing
%irreversible terms and the relaxation time, converging naturally to the
%ideal fluid case. This is not the case if we determine the irreversible
%currents by solving the differential equations such as Eq.(\ref{ISM}).
Finally, our simplified structure permits to include all of the irreversible
currents, such as bulk and shear viscosities and thermal conduction at the
same time without any difficulties.

The essential difference of our formalism from the IS is the expression for
entropy production. In the IS, the entropy production is required to be
positive definite algebraically. Thus, the integral form like our
formulation is not possible even neglecting some coupling terms. We relaxed
this condition, that is, the positiveness of entropy production is required
only for the hydrodynamical motion with time scales longer than the
relaxation time. For extremely violent change of variables, the
instantaneous entropy for a hydrodynamic cell would not necessarily be
positive definite.

We have applied our theory to the case of the one-dimensional scaling
solution of Bjorken and obtained the analogous behavior of previous
analysis. In this case we can prove explicitly the positiveness of entropy
production. We showed the time evolution of the temperature. As expected,
our theory gives the same result of Ref. \cite{Baier1}, because the
no-acceleration condition used in Ref. \cite{Baier1} is automatically
satisfied in this model. Note that our theory is applicable to more general
case where the acceleration is important.

The transport coefficients contained in relativistic dissipative
hydrodynamics should in principle be calculated from QCD. In the first order
theory, it is known that the transports coefficients can be calculated by
the Kubo formula \cite{Defu}. However, this formula does not gives the
transport coefficients in the second order theory, as is shown in Ref. \cite%
{Koide1} explicitly and the corresponding corrections should be evaluated.

Our theory is particularly adequate to be applied to the hydro-code such as
SPheRIO which is based on the Lagrangian coordinate system \cite%
{Spherio,Review}. Implementation of the present theory to the full
three-dimensional hydrodynamics is now in progress.

Authors are grateful for fruitful discussion of C. E. Aguiar, E.S. Fraga, T.
Hirano and T. Hatsuda. T. Koide would like to thank David Jou for useful
comments. This work was partially supported by FAPERJ, FAPESP, CNPq and
CAPES.

\end{document}